\documentclass{aastex}
\usepackage{spr-astr-addons}
\usepackage{url}

\RequirePackage{color}

\begin{document}

\title{  Penrose Process in Kerr-Taub-NUT Spacetime}
\shorttitle{Energetics in NUT Spacetime} \shortauthors{Shaymatov
et al.}

\author{A.A. Abdujabbarov\altaffilmark{1,2}} \email{ahmadjon@astrin.uz}  \and
\author{B.J. Ahmedov\altaffilmark{1,2,3}} \and
\author{S.R. Shaymatov\altaffilmark{1,2}}  \and
\author{A.S. Rakhmatov\altaffilmark{4}}

\altaffiltext{1}{Institute of Nuclear Physics,
        Ulughbek, Tashkent 100214, Uzbekistan}
\altaffiltext{2}{Ulugh Begh Astronomical Institute,
Astronomicheskaya 33, Tashkent 100052, Uzbekistan}
\altaffiltext{3}{International Centre for Theoretical Physics,
Strada Costiera 11, 34014 Trieste, Italy}
\altaffiltext{4}{National University of Uzbekistan,  Tashkent
100174, Uzbekistan}

\begin{abstract}

Penrose process on rotational energy extraction of the black hole
in the Kerr-Taub-NUT spacetime is studied. It has been shown that
for the radial motion of particles NUT parameter slightly shifts
the shape of the effective potential down. The dependence of the
extracted energy from compact object on NUT parameter has been
found.

\end{abstract}

\section{Introduction}\label{intro}

At present there is no any observational evidence for the
existence of gravitomagnetic monopole, i.e. so-called NUT (Newman,
Unti $\&$ Tamburino -- ~\citet{nut63}) parameter or {\it magnetic
mass}. Therefore study of the motion of the test particles and
energy extraction mechanisms in NUT spacetime may provide new tool
for studying new important general relativistic effects which are
associated with nondiagonal components of the metric tensor and
have no Newtonian analogues \citep[See, e.g.][where solutions for
electromagnetic waves and interferometry in spacetime with NUT
parameter have been studied.]{zonoz07,kkl08,ma08}. Kerr-Taub-NUT
spacetime with Maxwell and dilation fields is recently
investigated by~\citet{aliev08}. In our preceding
papers~\citep{mak08,aak08} we have studied the plasma
magnetosphere around a rotating, magnetized neutron star and
charged particle motion around compact objects immersed in
external magnetic field in the presence of the NUT parameter.

Penrose process~\citep[see, e.g.][]{pnr74} for the extraction of
energy from rotating black hole is based on the existence of
negative energy orbits in the ergosphere, the region bounded by
the event horizon and the static limit. \cite{gar2010} have
examined the possibility that astrophysical jet collimation may
arise from the geometry of rotating black holes and the presence
of high-energy particles resulting from a Penrose process rather
than from effect of  the magnetic fields. Detailed study of the
energetics of the Kerr-Newman black hole by the Penrose process is
given by~\cite{dadich84}. Energetics of a rotating charged black
hole in 5-dimensional supergravity has been recently considered
by~\cite{dadhich10}.

The geodesics  of the Kerr-Taub-NUT spacetime share many of the
properties of the ergosphere in the field of a magnetic monopole.
A thorough discussion and comparison of these orbits can be found
in papers~\citep{Zim89,Kax10}.

Here we first study motion of the test particles around rotating
compact object with nonvanishing NUT parameter. The effective
potential of the radial motion of the test particles around
rotating compact object is investigated in the presence of the NUT
parameter using the Lagrangian formalism. Finally we study the
energy extraction mechanism from compact object through Penrose
process in the Kerr-Taub-NUT spacetime.

The outline of the paper is as follow. In the Sec.~\ref{potential}
we study the ergosphere and motion of the test particles in the
Kerr-Taub-NUT spacetime. The Sec.~\ref{enextract} is devoted to
study the energy extraction mechanisms through Penrose process in
the Kerr-Taub-NUT spacetime. The concluding remarks are given in
the Sec.~\ref{conclusions}.

Throughout the paper, we use a space-like signature $(-,+,+,+)$
and a system of units in which $G = 1 = c$ (However, for those
expressions with an astrophysical application we have written the
speed of light explicitly.). Greek indices are taken to run from 0
to 3 and Latin indices from 1 to 3; covariant derivatives are
denoted with a semi-colon and partial derivatives with a comma.

\section{Ergosphere around compact object in Kerr-Taub-NUT spacetime}
\label{potential}

We consider electromagnetic fields of compact astrophysical
objects in Keer-Taub-NUT spacetime which in a spherical
coordinates $(c t,r,\theta,\phi )$ is described by the metric
\citep[see][]{Dadhich02,bini03}
\begin{eqnarray}\label{metric}
ds^{2} &=&-\frac{1}{\Sigma} \left(\Delta-a^{2}\sin^{2}
\theta\right)dt^{2}+\frac{\Sigma}{\Delta}dr^{2}+\Sigma
d\theta^{2} \nonumber\\
&&+ \frac{2}{\Sigma}\left[\Delta\chi
-a(\Sigma+a\chi)\sin^{2}\theta\right]dtd\varphi\nonumber\\
&&+\frac{1}{\Sigma}
\left[(\Sigma+a\chi)^{2}\sin^{2}\theta-\chi^{2}\Delta\right]
d\varphi^{2},
\end{eqnarray}
where parameters $\Sigma$, $\Delta$ and $\chi$ are defined by
\begin{eqnarray}
&&\Sigma=\emph{r}^{2}+(\emph{l}+a\cos\theta)^{2},\nonumber\\ &&
\Delta=\emph{r}^{2}-2\emph{Mr}-\emph{l}^{2}+a^{2},\nonumber\\
 &&\chi=a\sin^{2}\theta-2\emph{l}\cos\theta, \nonumber
\end{eqnarray}
$M$ is the total mass, $a$ is the specific angular momentum, and
$l$ is the NUT parameter of the central object.

The spacetime (\ref{metric}) has a horizon where the 4 velocity of
a co-rotating observer turns to null, or the surface $r=const$
becomes null:
\begin{eqnarray}
&& r_{+}= M + \sqrt{M^{2} + l^{2} - a^{2}}.
\end{eqnarray}

The static limit is defined where the time-translation Killing
vector $\xi_{(t)}\equiv\partial/\partial t$ becomes null (i.e.
$g_{00}=0$) and static limit of the black hole can be described as
\begin{eqnarray}
&&r_{st}= M + \sqrt{M^{2} + l^{2} - a^{2}\cos^{2}\theta }.
\end{eqnarray}

Considering only the outer horizon, $r_{+}$ and static limit, $
r_{\rm st} $, it can be verified that the static limit always lies
outside the horizon. The region between the two is called the
ergosphere, where timelike geodesics cannot remain static but can
remain stationary due to corotation  with the BH with the specific
frame dragging angular velocity at the given location in the
ergosphere. This is the region of spacetime where timelike
particles with negative angular momentum relative to the BH can
have negative energy relative to the infinity.

\begin{figure*}
a)  \includegraphics[width=0.45\textwidth]{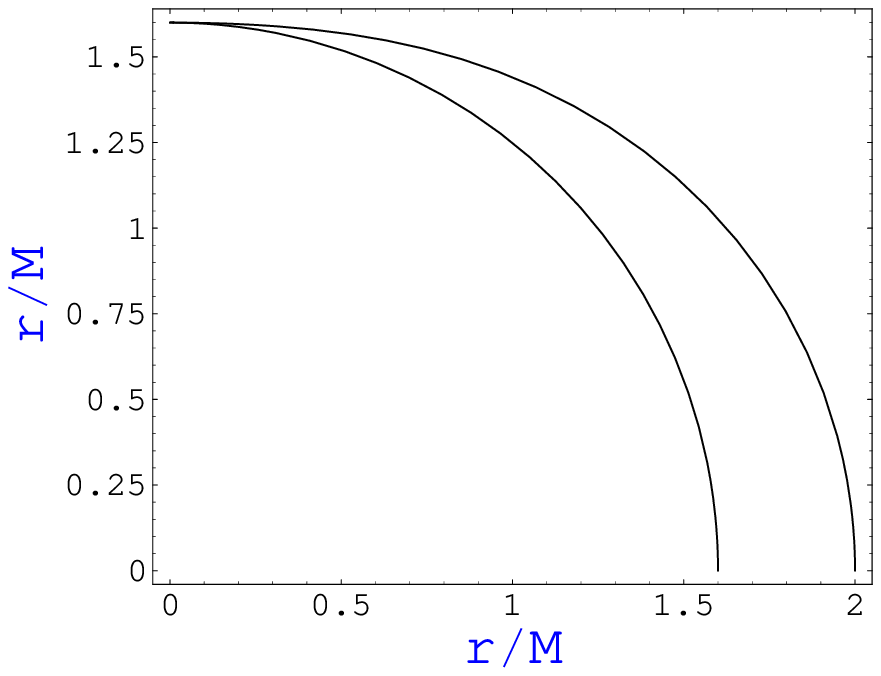}
b)  \includegraphics[width=0.45\textwidth]{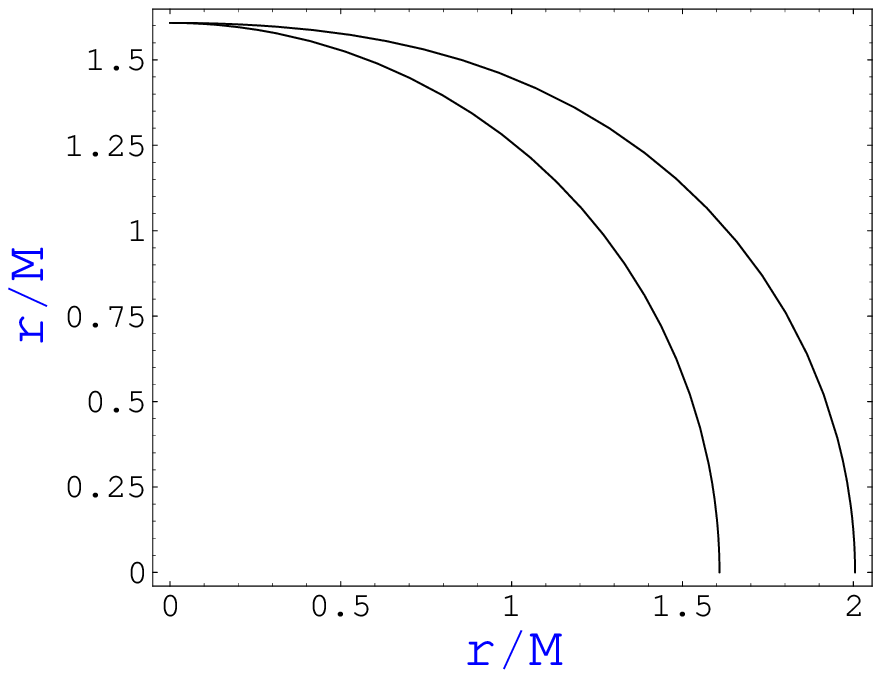} %

c)  \includegraphics[width=0.45\textwidth]{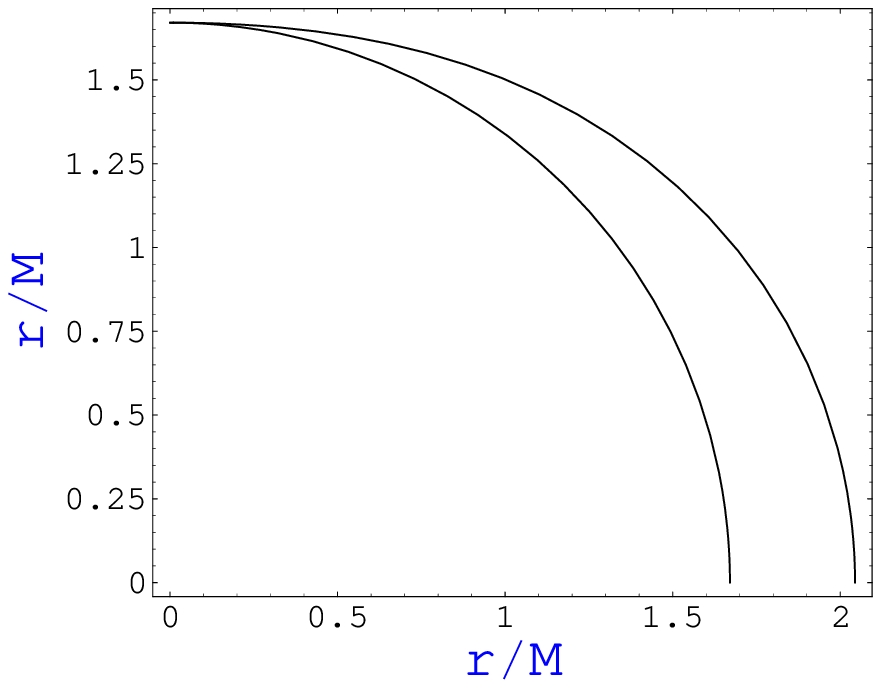} %
d)  \includegraphics[width=0.45\textwidth]{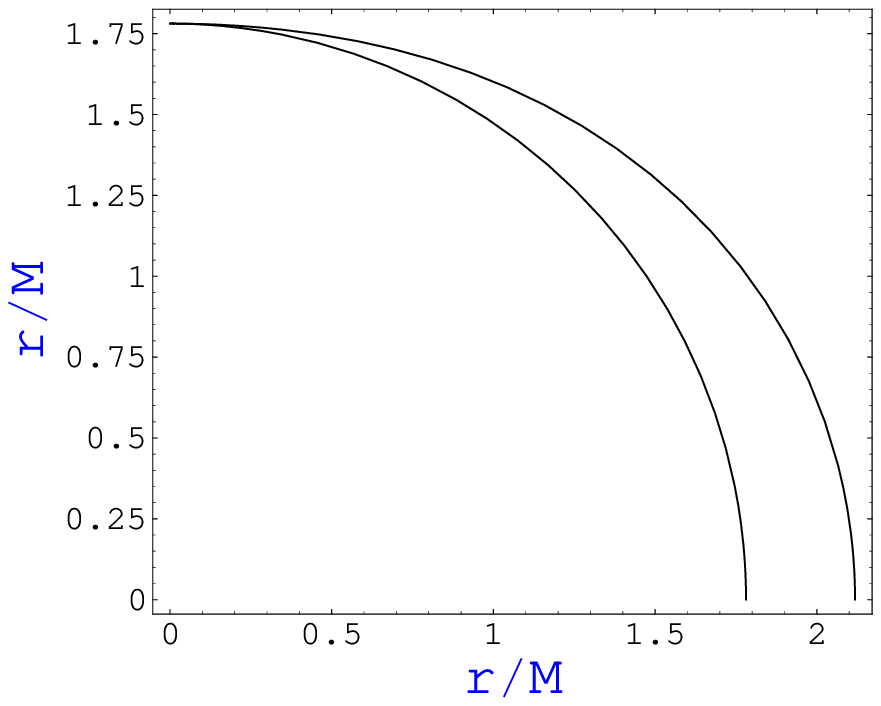} %
\caption{\label{ergosphere}The dependence of the shape of the
ergosphere from the small dimensionless NUT parameter $\tilde{l}$:
a) $\tilde{l}=0$, b) $\tilde{l}=0.1$ c) $\tilde{l}=0.3$ d)
$\tilde{l}=0.5$ .}
\end{figure*}

In Fig.\ref{ergosphere} the dependence of the shape of the
ergosphere from small dimensionless  parameter $\tilde{l}=l/M$ is
shown. From the dependence one can easily see that in the presence
of the NUT parameter radius of the event horizon becomes larger.
However a relative volume of the ergosphere is decreased.

Due to  the existence of an ergosphere around the black hole, it
is possible to extract energy from it by means of the Penrose
process. Inside the ergosphere, it is possible to have a timelike
or null trajectory with negative total energy. As a result, shoot
a small particle $A$ into the ergosphere from outside with energy
at infinity. When the particle is deep down near the horizon, let
it to explode into two parts, $B$ and $C$, one of which attains
negative energy relative to infinity and falls down the hole but
other part escapes back to radial infinity by conservation of
energy with energy greater than that of the original incident
particle. This is how the energy could be extracted from the hole
by axial accretion of particles with the suitable angular momentum
and $l$ parameters. Consider the equation of motion of such
negative energy particle. The Lagrangian for this particle can be
written as:
\begin{eqnarray}
2{\cal L}&=&-\frac{1}{\Sigma} \left(\Delta-a^{2}\sin^{2}
\theta\right)\dot{t}^{2}+\frac{\Sigma}{\Delta}\dot{r}^{2}+\Sigma
\dot{\theta}^{2} \nonumber\\
&&+ \frac{2}{\Sigma}\left[\Delta\chi
-a(\Sigma+a\chi)\sin^{2}\theta\right]\dot{t} \dot{\varphi}\nonumber\\
&&+\frac{1}{\Sigma}
\left[(\Sigma+a\chi)^{2}\sin^{2}\theta-\chi^{2}\Delta\right]
\dot{\varphi}^{2}, \label{1stcomp}
\end{eqnarray}
 and according to (\ref{1stcomp}) the generalized momenta are given by
\begin{eqnarray}
\label{E}  p_{t}&=&- \frac{1}{\Sigma}\left( \Delta-a^{2}\sin^{2}
\theta \right)
\dot{t}\nonumber\\
&&+\frac{1}{\Sigma}\left( \Delta\chi
-a(\Sigma+a\chi)\sin^{2}\theta  \right)
\dot{\varphi}= E ,\\
\label{L}
-p_{\varphi}&=& -\frac{1}{\Sigma} \left(\Delta\chi
-a(\Sigma+a\chi)\sin^{2}\theta \right)\dot{t}\nonumber\\
&&-\frac{1}{\Sigma}\left((\Sigma+a\chi)^{2}\sin^{2}\theta-\chi^{2}\Delta\right)\dot{\varphi}=L,\\
\label{P_{r}}
-p_{r}&=& -\frac{\Sigma}{\Delta}\dot{r},\\
\label{P_{teta}}
-p_{\theta}&=&-\Sigma \dot{\theta},
\end{eqnarray}
where superior dots denote differentiation with respect to an
affine parameter~$\tau$. (The conservation of $p_{t}$ and
$p_{\varphi}$ follows from the independence of the Lagrangian
 on  $\emph{t}$ and $\varphi$ which, in turn, is a manifestation of
 the stationary and the axisymmetric character of the Kerr-
 Taub-NUT geometry.)

The Hamiltonian for the test particle in spacetime (\ref{metric})
is given by
\begin{eqnarray}
{\cal H}&=&p_{t}\dot{t}+p_{\varphi}\dot{\varphi}+p_{r}\dot{r}+p_{\theta}\dot{\theta}-{\cal L} \nonumber\\
&&-\frac{1}{2 \Sigma} \left(\Delta-a^{2}\sin^{2}
\theta\right)\dot{t}^{2}+\frac{\Sigma}{2
\Delta}\dot{r}^{2}+\frac{1}{2}\Sigma
\dot{\theta}^{2} \nonumber\\
&&+ \frac{1}{\Sigma}\left[\Delta\chi
-a(\Sigma+a\chi)\sin^{2}\theta\right]\dot{t} \dot{\varphi}\nonumber\\
&&+\frac{1}{2 \Sigma}
\left[(\Sigma+a\chi)^{2}\sin^{2}\theta-\chi^{2}\Delta\right]
\dot{\varphi}^{2},
\end{eqnarray}
and from the independence of the Hamiltonian on $\emph{t}$ , one
can get
\begin{eqnarray}\label{H}
2{\cal H}  &=& \frac{\Sigma}{\Delta}\dot{r}^{2}+\Sigma
\dot{\theta}^{2}+\left[- \frac{1}{\Sigma}\left(
\Delta-a^{2}\sin^{2} \theta \right)
\dot{t}\right.\nonumber\\
&&\left.+\frac{1}{\Sigma}\left( \Delta\chi
-a(\Sigma+a\chi)\sin^{2}\theta  \right) \dot{\varphi}\right]
\dot{t}\nonumber\\
&& + \left[\frac{1}{\Sigma} \left(\Delta\chi
-a(\Sigma+a\chi)\sin^{2}\theta \right)\dot{t}\right.\nonumber\\
&&\left.+\frac{1}{\Sigma}\left((\Sigma+a\chi)^{2}\sin^{2}\theta-\chi^{2}\Delta\right)\dot{\varphi}\right]
\dot{\varphi}
\nonumber\\
&=&
E\dot{t}-L\dot{\varphi}+\frac{\Sigma}{\Delta}\dot{r}^{2}+\Sigma
\dot{\theta}^{2}=\delta=const.
\end{eqnarray}
We may, without loss of generality, set
\begin{eqnarray}
&&\delta=\left\{\begin{array}{ll}
1 & \mbox{~ for~ time-like ~geodesics}, \\
0 & \mbox{~ for~ null~ geodesics.}
\end{array}\right.
\end{eqnarray}
Then from the expressions (\ref{E}) and (\ref{L}) one can easily
obtain the expressions for $\dot{\varphi}$ and $\dot{t}$ as
\begin{eqnarray}\label{fdot}
\dot{\varphi}&=& \frac{1}{\Delta} \left[\frac{(\Delta -
a^{2}\sin^{2}\theta)(\chi E - L) - a \Sigma E
\sin^{2}\theta}{\Sigma \sin^{2}\theta}\right],\nonumber \\ \\
\label{tdot}
\dot{t}&=&-\frac{1}{\Delta} \left[ \frac{\Delta\Sigma
L}{\Delta\chi-a(\Sigma+a\chi)\sin^{2}\theta}\right.\nonumber\\
&&+\frac{(\Sigma+
a\chi)^{2}\sin^{2}\theta-\chi^{2}\Delta}{\Sigma\left(\Delta\chi-
a(\Sigma+a\chi)\sin^{2}\theta\right)}
\nonumber\\
&&\left.\times \left((\Delta-a^{2}\sin^{2}\theta)(\chi E-L)-
a\Sigma E\sin^{2}\theta\right)\right].\nonumber\\
\end{eqnarray}

{It was first shown by \citet{Zim89} for spherical symmetric case
(NUT spacetime) and later by \citet{bini03} for axial symmetric
case (Kerr-Taub-NUT spacetime) that the orbits of the test
particles are confined to a cone with the opening angle $\theta$
given by $\cos\theta = 2El/L$. It also follows that in this case
the equations of motion on the cone depend on $l$ only via $l^2$
\citep{bini03,aak08}. The main point is that the small value for
the upper limit for gravitomagnetic moment has been obtained by
comparing theoretical results with experimental data as (i) $l\leq
10^{-24}$ from the gravitational microlensing \citep{habibi04},
(ii) $l\leq 1.5\cdot10^{-18}$ from the interferometry experiments
on ultra-cold atoms \citep{ma08}, (iii) and similar limit has been
obtained from the experiments on Mach-Zehnder interferometer
\citep{kkl08}. Due to the smallness of the gravitomagnetic charge
let us consider the motion in the quasi-equatorial plane when the
motion in $\theta$ direction changes as
$\theta=\pi/2+\delta\theta(t)$, where $\delta\theta(t)$ is the
term of first order in $l$, then it is easy to expand the
trigonometric functions as $\sin\theta=1-\delta\theta^2(t)/2+{\cal
O}(\delta\theta^4(t))$ and $\cos \theta=\delta\theta(t)-{\cal
O}(\delta\theta^3(t))$.} Now inserting the expressions
(\ref{fdot}) and (\ref{tdot}) to the equation (\ref{H}), {and
neglecting the small terms ${\cal O}(\delta\theta^2(t))$,} one can
easily obtain equation for the radial motion as follows:
\begin{eqnarray}\label{radial}
\Sigma\dot{r}^{2} &=& 2 (L-aE)^{2}\frac{(Mr+l^{2})}{r^{2}+l^{2}}
\nonumber\\
&&+E^{2}(r^{2}+a^{2}+l^{2}) -L^{2}-\delta\Delta .
\end{eqnarray}
As we have noted, $\delta=0$ for null geodesics and equation for
radial motion  (\ref{radial}) becomes
\begin{equation}
\label{radial1} \Sigma\dot{r}^{2}=
2(L-aE)^{2}\frac{(Mr+l^{2})}{r^{2}+l^{2}}
+E^{2}(r^{2}+a^{2}+l^{2}) -L^{2} .
\end{equation}
Hereafter, it is more convenient to distinguish the geodesics by
the impact parameter $D={L}/{E}$. First, we study the geodesics with the impact parameter
\begin{eqnarray}
&& D=a, \mbox{~~when~~ } L=aE,
\end{eqnarray}
and equations (\ref{fdot}), (\ref{tdot}) and (\ref{radial1})
reduce to
\begin{eqnarray}
&& \dot{r}=\pm E,  \\
&& \dot{t}=-(r^{2}+a^{2}+l^{2})\frac{E}{\Delta},\\
&& \dot{\varphi}=-\frac{aE}{\Delta}.
\end{eqnarray}
The radial coordinate is uniformly described with respect to the
affine parameter while the equations governing $t$ and $\varphi$
are
\begin{eqnarray}
&& \frac{dt}{dr}=\pm \frac{r^{2}+a^{2}+l^{2}}{\Delta}, \\
&& \frac{d\varphi}{dr}=\pm \frac{a}{\Delta}.
\end{eqnarray}

\section{Energy extraction by Penrose process for Kerr-Taub-NUT
spacetime}
\label{enextract}

{Let us continue our assumption that the deflection in $\theta$
direction is reasonably small and orbits of the particles are in
the quasi-equatorial plane $\theta=\pi/2+\delta\theta(t)$. Using
the assumptions mentioned in previous section one can easily
rewrite the expressions (\ref{E}) and (\ref{L}) in the
approximation ${\cal O}(\delta\theta^2(t))$ as a quadratic
equation in energy:}
\begin{eqnarray}\label{EnergyF}
&&\alpha E^{2}-2\beta E+\gamma+\frac{\Sigma}{\Delta}(p^{r})^{2}+
\Sigma(p^{\theta})^{2}+ m^{2}=0,\nonumber\\
\end{eqnarray}
where we have used the following notations
\begin{eqnarray}
&&\alpha=-\left(1-2\frac{Mr+l^{2}}{r^{2}+l^{2}}\right),\\
&& \beta=4a\frac{Mr+l^{2}}{r^{2}+l^{2}}L, \\
&&\gamma=\left(r^{2}+a^{2}+l^{2}+2a^{2}\frac{Mr+l^{2}}{r^{2}+l^{2}}\right)L^{2}.
\end{eqnarray}
From the equation (\ref{EnergyF}) one can easily obtain the
equation of radial motion in the following form:
\begin{eqnarray}
&& \dot{r}^2=E^2-V_{\rm eff}\ ,
\end{eqnarray}
and the notation
\begin{eqnarray}
&& V_{\rm eff}= E^{2}-2\frac{(M r + l^{2})(L - a E)^{2}}{\Sigma
(r^{2} + l^{2})}\nonumber \\
&&\qquad - \frac{E^{2}(r^{2} + a^{2} + l^{2})}{\Sigma} +
\frac{L^{2} + \Delta}{\Sigma}
\end{eqnarray}
denotes the effective potential of the radial motion of the test
particle around rotating compact object with nonvanishing NUT
parameter.
\begin{figure}

\includegraphics[width=0.45\textwidth]{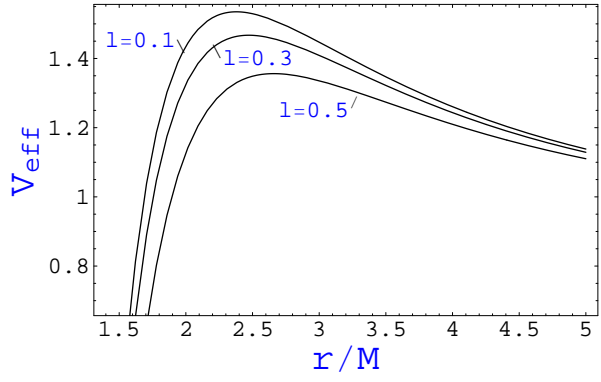}

\caption{\label{effpot1}The radial dependence of the effective
potential of radial motion of the particle for the different
values of the dimensionless NUT parameter $\tilde{l}$ .}
\end{figure}

In the Fig. \ref{effpot1} the radial dependence of the effective
potential of  radial motion of the massive test particle has been
shown for the different values of the dimensionless parameter
$\tilde{l}$. Here for the energy and momenta of the particle the
following values are taken: $E/m=0.9$, $L/mM=4.3$. The presence of
the parameter $l$ slightly shifts the shape of the effective
potential down.

Now, as the particle falls through the horizon, the mass of the
black hole will change by $\delta M=E$. {There is no upper limit
for change of mass of the central black hole. The infalling big
number of particles with positive energy can essentially increase
the mass of the black hole. }
But there is a lower limit on $\delta M$ which
could be added to the black hole corresponding to $m=0$,
$p^{\theta}=0$ and $p^{r}=0$. Evaluating all of the required
quantities at the horizon $r=r_{+}$, we get the limit for the
change in black hole mass as

\begin{eqnarray}
&&\delta M=-\frac{L_{z}a(Mr_{+}+l^{2})}{r^{2}_{+}-2Mr_{+}}D ,
\end{eqnarray}
where
\begin{eqnarray}
D&=& 4\bigg[ 1 - \bigg\{1+
{\left(1-\frac{a^{2}(r^{2}_{+}+2Mr_{+}+3l^{2})}{
(r^{2}_{+}+l^{2})^{2}}\right)}\nonumber
\\
&& \frac{\left(r_{+}^{2}- 2Mr_{+}-l^{2}\right)}{16a^{2}}
{\left(\frac{Mr_{+}+l^{2}}{r^{2}_{+}+l^{2}}
\right)^{-2}}\bigg\}^{\frac{1}{2}}\bigg].
\end{eqnarray}

\begin{figure}

\includegraphics[width=0.45\textwidth]{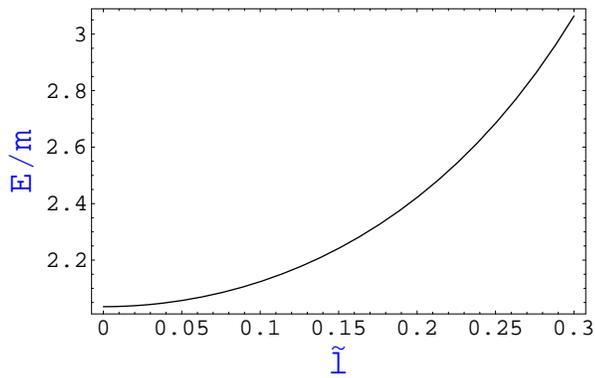}

\caption{\label{fig:3}The dependence of the extracted energy from
the black hole on the dimensionless NUT parameter $\tilde{l}$ . }

\end{figure}

To be able to extract energy from the black hole $(\delta M<0)$,
we must therefore have
\begin{eqnarray}
&&\frac{L_{z}a(Mr_{+}+l^{2})}{r^{2}_{+}-2Mr_{+}}D>0.
\end{eqnarray}
In the Fig. \ref{fig:3} the dependence of the extracted energy
from the black hole on the small dimensionless parameter
$\tilde{l}$ has been shown: with increasing the parameter
$\tilde{l}$ the relative extraction of the energy becomes more
stronger. The graph shows that the extraction of the energy is
directly proportional to the parameter $\tilde{l}$ , that is with
increasing $\tilde{l}$, the extracted energy also increases.

\section{Conclusion}
\label{conclusions}

We have studied  the properties of the ergosphere of the black
hole in the Kerr-Taub-NUT spacetime. The dependence of the shape
of the ergosphere from small dimensionless NUT parameter shows
that the radius of the event horizon becomes larger. However the
relative volume of the ergosphere is decreased and the extracted
energy from the black hole is raised up by the small dimensionless
parameter $\tilde{l}$ .

\section*{Acknowledgments}


This research is supported in part by the UzFFR (projects 1-10 and
11-10) and projects FA-F2-F079, FA-F2-F061 of the UzAS and by the
ICTP through the OEA-PRJ-29 project.

\end{document}